\documentclass[prd,preprintnumbers,showkeys,nofootinbib,12pt]{article}
\usepackage{epsfig}

\usepackage{bm, graphicx,amsmath,amssymb,epsf}
\usepackage{appendix}

\usepackage{verbatim}

\usepackage{subfigure}

\usepackage{graphicx}
\newcommand{\beq}{\begin{equation}}
\newcommand{\eeq}{\end{equation}}
\newcommand{\beqn}{\begin{eqnarray}}
\newcommand{\eeqn}{\end{eqnarray}}
\newcommand{\beqs}{\begin{eqnarray*}}
\newcommand{\eeqs}{\end{eqnarray*}}

\begin{document}

\title{\large \bf Mandelstam cuts and light-like Wilson loops in $\mathcal{N}=4$  SUSY  }
\author{\large  L.~N. Lipatov$^{1,2}$ and  
A.~Prygarin$^{1}$ \bigskip \\
{\it 
$^1$~II. Institute of  Theoretical Physics, Hamburg University, Germany} \\
{\it  $^2$~St. Petersburg Nuclear Physics Institute, Russia}}

\maketitle

\vspace{-9cm}
\begin{flushright}
DESY-10-124
\end{flushright}
\vspace{8cm}

\abstract{We perform  an analytic continuation of  the two-loop remainder function for the six-point planar MHV amplitude in  $\mathcal{N}=4$ SUSY, found by Goncharov, Spradlin, Vergu and Volovich from the light-like Wilson loop representation. The remainder function is continued into a physical region, where all but two energy invariants are negative. It turns out to be pure imaginary in the multi-Regge kinematics, which is  in an agreement with the predictions based on the Steinmann relations for the Regge poles and Mandelstam cut contributions. The leading term reproduces correctly the expression calculated by one of the authors in the BFKL approach, while the subleading term presents a result, that was not yet found with the use of the unitarity techniques.   
This supports the applicability  of the Wilson loop approach to  the planar MHV 
amplitudes  in  $\mathcal{N}=4$ SUSY.   
}




\section{Introduction}

In  recent years a significant progress was reached in revealing the structure of scattering amplitudes in the supersymmetric theories. Parke and Taylor~\cite{Parke:1986gb}   first showed that so-called maximally helicity violating~(MHV) gluon scattering amplitudes at tree level can be written in a very compact way. This suggests that quantum corrections can be also included in a more  efficient way than in the framework of the traditional Feynman technique. A great effort in that direction led to formulation of ABDK~\cite{Anastasiou:2003kj} and then BDS~\cite{BDS} ansatz for multi-loop planar MHV amplitudes in SYM $\mathcal{N}=4$. The BDS formula was claimed to account for all loop corrections by exponentiation of the one loop result. However it was shown by one of the authors with collaborators~\cite{BLS1} that the BDS ansatz for six-point amplitude at two loops  is not compatible with Steinmann relations~\cite{Steinmann}, claiming the absence of simultaneous singularities in the overlapping channels.  

Drummond, Henn, Korchemsky and Sokatchev~\cite{Drummond:2007au} analyzing the conformal properties of polygon Wilson loops showed that anomalous conformal Ward identities  uniquely
fix the form of the all-loop $4$- and  $5$-point amplitudes, so that any relative correction to the BDS ansatz starting at six external points is necessarily a function of conformal invariants (cross ratios of dual coordinates).  
  The correction to the BDS formula was called the remainder function $R^{(L)}_{n}$ for an amplitude with $L$ loops and $n$ external legs, and the first non-trivial remainder function is  $R^{(2)}_{6}$.
The imaginary part of $R^{(2)}_{6}$ in the leading logarithm approximation~(LLA) of the BFKL approach~\cite{BFKL} was calculated by one of the authors with collaborators~\cite{BLS2}. For general $n$ the remainder function contains contributions of Mandelstam cuts constructed from an arbitrary number of reggeized gluons with the local Hamiltonian of an integrable Heisenberg spin chain~\cite{Lipatov:2009nt}.
 
 It was suggested~\cite{AldayMalda,DKS,Brandhuber:2007yx,Berkovits:2008ic,Drummond:2007bm,Drumlast} that $R^{(L)}_{n}$ can be obtained from the expectation value of the light-like two-loop hexagon Wilson loop in SYM $\mathcal{N}=4$.   Del Duca, Duhr and Smirnov~\cite{DelDuca:2009au,DelDuca:2010zg} expressed $R^{(2)}_{6}$ in terms of   generalized polylogarithms, which was greatly simplified by Goncharov, Spradlin, Vergu and Volovich~(GSVV) \cite{Goncharov:2010jf}, and written in terms of $\text{Li}_k$ functions only  with arguments depending on three  cross ratios $u_1$, $u_2$ and $u_3$. 

The objective of the present study is to compare analytically the remainder function $R^{(2)}_{6}$ calculated from the expectation value of the light-like hexagon Wilson loop~\cite{Goncharov:2010jf} and its  imaginary part found in the BFKL approach~\cite{BLS2}. Numerically, an agreement between the two approaches was demonstrated by Schabinger~\cite{Schabinger:2009bb}. The leading correction to $R^{(2)}_{6}$ coincides with the BFKL predictions and the next-to-leading term is  pure imaginary in an agreement with the expectations based on  analytic properties of the production amplitudes~\cite{Lipatov:NEW}.

In the next section we present a result of the analytic continuation of the GSVV formula into a  region of multi-Regge kinematics, where all but two energy invariants are negative. The rest of the paper is devoted to details of the analytic continuation and to the analysis  of the obtained result.   

\section{Analytic continuation of GSVV formula}

\hspace{0.3cm} The remainder function for six-point amplitude depends only of three cross ratios of dual coordinates in accordance to ref.~\cite{Drummond:2007au}.
These cross ratios can be expressed through the kinematic invariants shown in Fig.~\ref{fig:6point} as follows
\begin{eqnarray}\label{crossinv}
 u_1=\frac{s\;s_2}{s_{012}\;s_{123}},\;\;\; u_2=\frac{s_1\;t_3}{s_{012}\;t_{2}},\;\;\; u_3=\frac{s_3\;t_1}{s_{123}\;t_2}.
\end{eqnarray}

The GSVV~\cite{Goncharov:2010jf}  formula for the remainder function reads

\begin{eqnarray} \label{R6}
R(u_1,u_2,u_3) = \sum_{i=1}^3 \left( L_4(x^+_i, x^-_i) -
\frac{1}{2} \text{Li}_4(1 - 1/u_i)\right) \cr
- \frac{1}{8} \left( \sum_{i=1}^3 \text{Li}_2(1 - 1/u_i) \right)^2
+ \frac{J^4}{24} + \chi \frac{\pi^2}{12} \left(  J^2 +  \zeta(2)\right),
\end{eqnarray}
where 
\begin{equation}\label{Xpm}
x^\pm_i = u_i x^\pm, \qquad
x^\pm = \frac{u_1+u_2+u_3-1 \pm \sqrt{\Delta}}{2 u_1 u_2 u_3},
\end{equation}
and $\Delta = (u_1+u_2+u_3-1)^2 - 4 u_1u_2u_3$.

The function $L_4(x^+, x^-)$ is defined by 

\begin{equation}
\label{eq:bwrz}
L_4(x^+, x^-) = \sum_{m=0}^3
\frac{(-1)^m}{(2m)!!} \log(x^+ x^-)^m
(\ell_{4-m}(x^+) + \ell_{4-m}(x^-))
+ \frac{1}{8!!} \log(x^+ x^-)^4,
\end{equation}
together with
\begin{equation}
\ell_n(x) = \frac{1}{2} \left( \text{Li}_n(x) - (-1)^n \text{Li}_n(1/x) \right),
\end{equation}
as well as the quantities
\begin{equation}\label{J}
J = \sum_{i=1}^3 (\ell_1(x^+_i) - \ell_1(x^-_i)),
\end{equation}
and 
\begin{equation}\label{chi}
\begin{aligned}
\chi &= \begin{cases}
-2 & \Delta > 0 ~{\rm and}~ u_1+u_2+u_3>1,\cr
+1 & {\rm otherwise}.
\end{cases}
\end{aligned}
\end{equation}

The remainder  function $R(u_1,u_2,u_3)$ was found in  the region where all cross ratios $u_i$ are positive. 
The multi-Regge kinematics~(MRK) is defined by 
\beqn\label{MRKinv}
s \gg s_{012},s_{123}\gg s_1,s_2,s_3 \gg t_1,t_2,t_3 
\eeqn
for the kinematic invariants depicted in Fig.~\ref{fig:6point}. 
For the cross ratios of Eq.~\ref{crossinv} the multi Regge kinematics implies (cf.~\cite{BLS2}) 
 \begin{eqnarray}\label{MRK}
1-u_1\to +0,\;\; u_{2}\to +0,\;\;u_{3}\to + 0,\;\; \frac{u_2}{1-u_1}\simeq \mathcal{O}(1)
,\;\; \frac{u_3}{1-u_1}\simeq \mathcal{O}(1).
\end{eqnarray}
In this kinematics the remainder  function $R(u_1,u_2,u_3)$ vanishes until $u_1$ has a phase. It was argued in ref.~\cite{BLS1} that in some  physical channels for  the planar   six-point scattering amplitudes the variable $u_1$ can develop a phase of $\phi=\pm 2\pi$. This phase results into a non-vanishing pure imaginary   
contribution to $R(u_1,u_2,u_3)$ in the multi-Regge kinematics. The non-vanishing term of the remainder  function for $u_1=e^{-i2\pi}|u_1|$ originates from the Mandelstam cuts~\cite{Mandelstam}, which are not captured by the BDS formula. This contribution was found \cite{BLS2} in the leading logarithmic approximation~(LLA), which keeps only the terms leading in the logarithm of  energy~$\sqrt{s_2}$. 

In the present study we perform the  analytic continuation of Eq.~\ref{R6} into a region with  $u_1=e^{-i2\pi}|u_1|$ and then extract both, the leading and next-to-leading~(NLO) terms in the multi-Regge kinematics of  Eq.~\ref{MRK}. The result of this analytic continuation is rather compact and reads 
\begin{eqnarray}\label{KR62result}
&&R(|u_1|e^{-i2\pi},|z|^2(1-u_1),|1-z|^2(1-u_1))\simeq \frac{i\pi}{2}\ln(1-u_1)\ln |z|^2 \ln |1-z|^2  \nonumber\\
&&+\frac{i\pi}{2} \ln \left(|z|^2|1-z|^2\right) \left(\ln z \ln (1-z)+\ln z^* \ln (1-z^*) -2\zeta_2\right)   \nonumber\\
&& +\frac{i\pi}{2} \ln \frac{|1-z|^2}{|z|^2}\left(\text{Li}_2(z)+\text{Li}_2(z^*)-\text{Li}_2(1-z)-\text{Li}_2(1-z^*)\right)
\nonumber \\
&& +i2\pi \left(\text{Li}_3(z)+\text{Li}_3(z^*)+\text{Li}_3(1-z)+\text{Li}_3(1-z^*)-2\zeta_3\right).
\end{eqnarray}
In Eq.~\ref{KR62result} we introduced complex variables 
\begin{eqnarray}\label{Kcomplex}
z=\sqrt{\frac{u_2}{1-u_1}}e^{i\phi_2},\;\;\;1-z=\sqrt{\frac{u_3}{1-u_1}}e^{-i\phi_3}
\end{eqnarray}
to remove some square roots in the arguments of the polylogarithms~(see Eq.~\ref{Xpm}).
The parametrization of Eq.~\ref{Kcomplex} is compatible with the constraints of the multi-Regge kinematics as  discussed in the following sections. 

The term on the RHS of the first line of Eq.~\ref{KR62result} coincides with the LLA term found by one of the 
authors with collaborators~\cite{BLS2,Lipatov:NEW} using the BFKL approach. Other terms in Eq.~\ref{KR62result}, that are not accompanied by $\ln(1-u_1)$,  are subleading in the logarithm of energy $\sqrt{s_2}$ and were not yet calculated in the BFKL formalism.  The complex variable  $z$
does not depend on energy, and is a function of transverse momenta only as follows  from Eq.~\ref{Kcomplex},  Eq.~\ref{redcross} and  Eq.~\ref{redcrosstrans}.  
$R(|u_1|e^{-i2\pi},|z|^2(1-u_1),|1-z|^2(1-u_1))$ is pure imaginary, in full agreement with analyticity predictions~\cite{Lipatov:NEW}. It is also invariant under transformations $z \leftrightarrow 1-z$, which correspond to $u_2 \leftrightarrow u_3$ invariance, related to the target-projectile symmetry. Eq.~\ref{KR62result}  vanishes for $z\to 1$ or $z\to 0$, when the momentum of one of the produced particles $k_i$ in Fig.~\ref{fig:6point} goes to zero, in an accordance to the expectation that in the collinear limit the six-point amplitude is reduced to the five-point amplitude which does not contain Mandelstam cuts. 

This way we find an agreement 
between the Wilson loop result and the BFKL approach, at least at the level of the leading logarithm  approximation.
The expression in Eq.~\ref{KR62result} is the main result of the present study. Some details and discussions of the analytic continuation are presented in the following sections.

\section{BFKL approach and BDS amplitudes}

In this section we briefly outline the result of ref.~\cite{BLS2}, where the BDS violating  piece was found analytically in LLA.

A simple ansatz for gluon production amplitudes with the maximal helicity 
violation in a planar limit for SYM $\mathcal{N}=4$  was suggested by Bern, 
Dixon and Smirnov~\cite{BDS}.
But  it was shown in ref.~\cite{BLS1}, that for the 6-point case 
this ansatz is in a disagreement with the 
Steinmann relations~\cite{Steinmann} which are equivalent to the statement, that
the production amplitude in the physical regions should not have simultaneous
singularities in overlapping channels.
 The analogous conclusion about the violation of the BDS ansatz  was reached in the  numerical studies of the six-point amplitude at two loops~\cite{BernDrum}. The reason 
for the disagreement 
is related to the fact, that the BDS amplitude for the transition 
$2\rightarrow 4$ in
the multi-Regge kinematics does not contain in the
$j_2$-plane of the $t_2$ channel the Mandelstam cut contribution appearing 
in the physical kinematical regions, where the invariants in the direct channels
have the following signs $s,s_2>0;\,s_1,s_3<0$ or  $s,s_2<0;\,s_1,s_3>0$. In LLA
this contribution for the 6-point
amplitude was calculated with the use of the BFKL equation~\cite{BLS2}. The
corresponding amplitude in the region $s,s_2>0;\,s_1,s_3<0$
can be written in the factorized form
\beq
M_{2\rightarrow 4}=M^{BDS}_{2\rightarrow 4}\,(1+i\Delta _{2\rightarrow 4}),
\label{corLLA}
\eeq
where $A^{BDS}$ is the BDS amplitude~\cite{BDS} and the correction $\Delta _{2\rightarrow 4}$ was calculated 
in all orders with a logarithmic accuracy
\beq
i\Delta _{2\rightarrow 4}=\frac{a}{2}\, \sum _{n=-\infty}^\infty (-1)^n
\int _{-\infty}^\infty \frac{d\nu }{\nu ^2+\frac{n^2}{4}}\, 
\left(\frac{q_3^*k^*_1}{k^*_2q_1^*}\right)^{i\nu -\frac{n}{2}}\,
\left(\frac{q_3k_1}{k_2q_1}\right)^{i\nu +\frac{n}{2}}\,
\left(s_2^ {\omega (\nu , n)}-1\right)\,.
\label{LLA}
\eeq
Here $k_1,k_2$ are transverse components of produced gluon momenta,
$q_1,q_2,q_3$ are the momenta of reggeons in the corresponding
crossing channels and 
\beq
\omega (\nu , n)=4a\,
\Re \left(2\psi (1)-\psi (1+i\nu +\frac{n}{2})-\psi (1+i\nu -\frac{n}{2})\right).
\label{eigen}
\eeq
The LLA correction to the BDS formula in Eq.~\ref{LLA} is valid at any number of loops and, for example, is calculated explicitly at three loops in ref.~\cite{Prygarin:NEW}.

\begin{figure}[htbp]
	\begin{center}
	\subfigure[$u_1=e^{i0}|u_1|$~($s,s_2,s_1,s_3>0$)]{
	\epsfig{figure=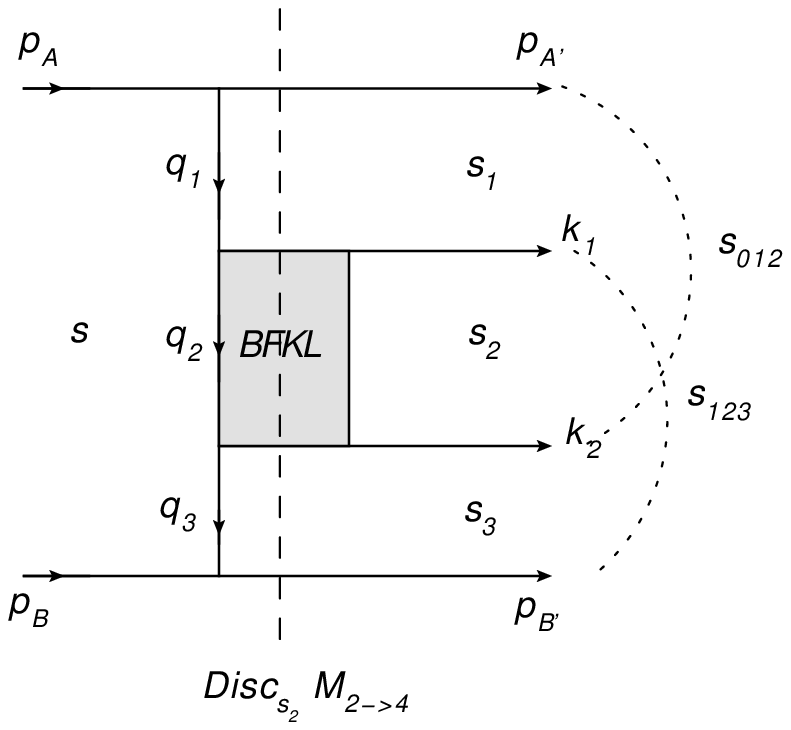,width=65mm}}
	\subfigure[$u_1=e^{-i2\pi}|u_1|$~($s,s_2>0;\,s_1,s_3<0$)]{
	\epsfig{figure=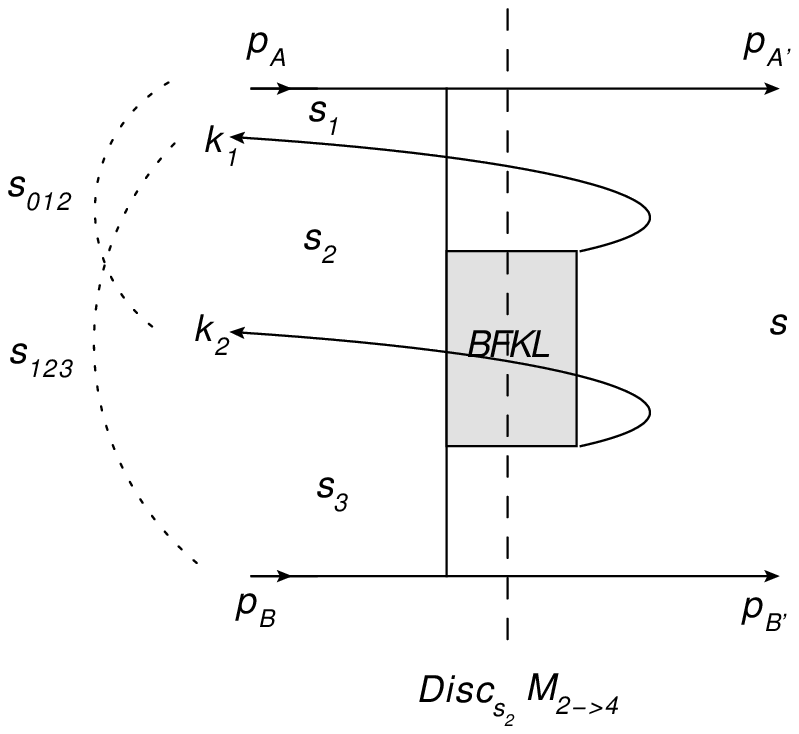,width=65mm}}
		\end{center}
	\caption{ The BDS violating part appears in the region $s,s_2>0;\,s_1,s_3<0$.   }
	\label{fig:6point}
\end{figure}
The correction $\Delta _{2\rightarrow 4}$ is M\"{o}bius invariant in the transverse momentum space and can be
written in terms of the four-dimensional anharmonic ratios~\cite{BLS2} in an accordance 
to the results of refs.~\cite{DKS,DrumSmir}. 
The corresponding $4$-dimensional cross ratios are expressed through the energy invariants (see Eq.~\ref{crossinv})  shown in Fig.~\ref{fig:6point}.  
 From Eq.~\ref{crossinv} we can calculate the phases of the cross ratios in the indirect channel depicted in Fig.~\ref{fig:6point}b with  respect to the cross ratios  of the direct channel  Fig.~\ref{fig:6point}a. When we flip the momenta of the produced particles $k_1$ and $k_2$, all, but $s$ and $s_2$, energy invariants $s_i$  change the sign (or multiplied by $e^{i\pi}$). The relative phases cancel in $u_2$ and $u_3$, but add up to $e^{-i2\pi}$ in the cross ratio $u_1$. So that according to Fig.~\ref{fig:6point}
 \begin{eqnarray}\label{u1cont}
 {u_1}_{b}=e^{-i2\pi}{u_1}_{a}. 
 \end{eqnarray}

For the purpose of the present discussion it is useful to introduce   reduced cross ratios
\begin{eqnarray}\label{redcross}
 \tilde{u}_2=\frac{u_2}{1-u_1}, \;\;\;  \tilde{u}_3=\frac{u_3}{1-u_1}.
\end{eqnarray}

Using the onshellness of the particles with momenta $k_1$ and $k_2$ one can show that the reduced  $4$-dimensional cross ratios $\tilde{u}_2$
and $\tilde{u}_3$  do not include $s$ or $s_2$ and depend only on the transverse $2$-dimensional  momenta in the multi-Regge kinematics 
\begin{eqnarray}\label{redcrosstrans}
 \tilde{u}_2=\frac{|\bold{k}_2|^2|\bold{q}_1|^2}{|\bold{k}_2+\bold{k}_1|^2 |\bold{q}_2|^2} , \;\;\;\tilde{u}_3=\frac{|\bold{k}_1|^2|\bold{q}_3|^2}{|\bold{k}_2+\bold{k}_1|^2 |\bold{q}_2|^2}.
\end{eqnarray}
Thus any function of  $\tilde{u}_2$
and $\tilde{u}_3$ only in the remainder function is subleading in the leading logarithm of the energy and corresponds to the NLO contributions.

The BDS violating piece in two loops  found in ref.~\cite{BLS2}  can be written in terms of the reduced cross ratios as 
\begin{eqnarray}\label{KblsR6}
 && i\Delta _{2\rightarrow 4}=-i2\pi \frac{a^2}{4}\ln s_2 \ln \left(\frac{|\bold{k}_2+\bold{k}_1|^2 |\bold{q}_2|^2}{|\bold{k}_2|^2|\bold{q}_1|^2}\right)
\ln \left(\frac{|\bold{k}_2+\bold{k}_1|^2 |\bold{q}_2|^2}{|\bold{k}_1|^2|\bold{q}_3|^2}\right)  \\
&&\simeq a^2\frac{i\pi}{2}\ln(1-u_1)\ln \tilde{u}_2 \ln \tilde{u}_3
\nonumber
\end{eqnarray}
using $1-u_1\simeq (\mathbf{k}_1+\mathbf{k}_2)^2/s_2$.
Eq.~\ref{KblsR6} follows also from general arguments related to the analyticity and factorization of the production amplitudes~\cite{Lipatov:2009nt,Lipatov:NEW}.
With the help of  Eq.~\ref{corLLA} this recasts in  a  form of the two-loop correction to the BDS formula
\beq
M^{(2)}_{2\rightarrow 4}=M^{(2)BDS}_{2\rightarrow 4}\,\left(1+a^2\frac{i\pi}{2}\ln(1-u_1)\ln \tilde{u}_2 \ln \tilde{u}_3\right).
\label{corLLAreduced}
\eeq
In  the complex variables of Eq.~\ref{Kcomplex} it reads

\beqn
M^{(2)}_{2\rightarrow 4}=M^{(2)BDS}_{2\rightarrow 4}\,\left(1+a^2\frac{i\pi}{2}\ln(1-u_1)\ln |z|^2 \ln |1-z|^2\right).
\label{KcorLLAreduced}
\eeqn

The BDS violating LLA term found in the BFKL approach and given by Eq.~\ref{KcorLLAreduced} is reproduced by the Wilson loop result in Eq.~\ref{R6} after its  analytic continuation as one can see from Eq.~\ref{KR62result}.

\section{Analytic continuation -discussions}

In this section we perform an analytic continuation of the GSVV formula in Eq.~\ref{R6} for the remainder  function and then extract the leading and subleading terms in the   multi-Regge kinematics~(see Eq.~\ref{KR62result}). In terms of the cross ratios this kinematics corresponds to the limit  Eq.~\ref{MRK} as can be seen from the energy dependence of the cross ratios in  Eq.~\ref{crossinv}. Eq.~\ref{MRK} describes both the direct (see Fig.~\ref{fig:6point}a) and the indirect (see Fig.~\ref{fig:6point}b) channels.  The GSVV formula is valid in the region, where all cross ratios are positive $u_i>0$. It is real valued and vanishes in the direct channel in the limit Eq.~\ref{MRK}. However, in the course of   the analytic continuation the function  $R(u_1,u_2,u_3)$ in Eq.~\ref{R6} may develop an imaginary part, which does not vanish in the multi-Regge kinematics.  This non-vanishing contribution is related to the presence of the Mandelstam cut, which is not captured by the BDS ansatz. In the BFKL approach this  cut is described by the propagation of a  color octet object  built of two reggeized gluons and referred to as the composite color  BFKL state. According to Eq.~\ref{u1cont}, the analytic continuation from the direct channel to the indirect channel is performed continuing the remainder  function along the circle 
 \begin{eqnarray}\label{u1phi}
 u_1=e^{i\phi}|u_1|
 \end{eqnarray}
 from $\phi=0$ to $\phi=-i2\pi$. Other two cross ratios $u_2$ and $u_3$ remain untouched, because they are the same both in the direct and the indirect channels. A certain care should  be taken when continuing Eq.~\ref{R6} because of the definition of the function $\chi$ in Eq.~\ref{chi} as a step function that can potentially cause problems on the boundary of $u_1+u_2+u_3>1$ and $\Delta>0$. To avoid this difficulty we pick up a region where $u_1+u_2+u_3<1$   for  $\phi=0,-i\pi,-i2\pi$ in Eq.~\ref{u1phi} and thus $\chi$ does not change its value $\chi=1$ during the analytic continuation at these points.  It is worth emphasizing that the function $\Delta$ does change its sign  for $\phi=-i\pi$, but this does not affect the value of $\chi$ since the condition $u_1+u_2+u_3>1$  in Eq.~\ref{chi} is  not fulfilled.
 
\begin{figure}[htbp]
	\begin{center}
		\epsfig{figure=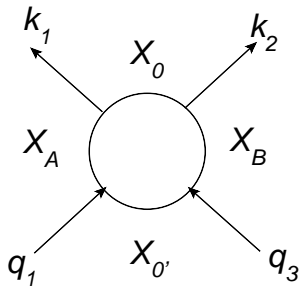,width=40mm}
	\end{center}
	\caption{The dual coordinates of the transverse momenta.   }
	\label{fig:dual}
\end{figure}

A few words to be said about the constraints on the anharmonic ratios $u_i$ set by the multi-Regge kinematics in the physical region.
It is convenient to pass to the dual coordinates~(see  Fig.~\ref{fig:dual}) of the transverse momenta for the reduced cross ratios in Eq.~\ref{redcrosstrans}.

\begin{figure}[htbp]
	\begin{center}
		\epsfig{figure=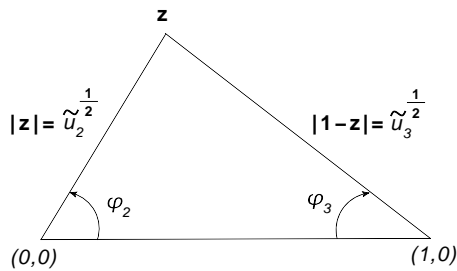,width=60mm}
	\end{center}
	\caption{The "unitarity" triangle.   }
	\label{fig:triangle}
\end{figure}

In terms of the dual coordinates the reduced cross ratios in Eq.~\ref{redcrosstrans} read
\begin{eqnarray}\label{crossX}
\tilde{u}_2=\frac{|x_{0B}|^2|x_{0'A}|^2}{|x_{AB}|^2|x_{00'}|^2}, \;\;\;
 \tilde{u}_3=\frac{|x_{0A}|^2|x_{0'B}|^2}{|x_{AB}|^2|x_{00'}|^2}.
\end{eqnarray}
Let us find  restrictions on the cross ratios imposed by the multi-Regge kinematics.
Due to the M\"obius invariance we can put 
\beq
x_{A}=1,\;\;\; x_{B}=0,\;\;\; x_{0'}=\infty,\;\;x_{0}=z,
\eeq 
then 
\beq\label{Kvarcomplex}
\tilde{u}_2=|z|^2, \;\;\; \tilde{u}_3=|1-z|^2.
\eeq
So that the reduced crossed ratios are related to the "unitarity" triangle  as depicted in Fig.~\ref{fig:triangle}.
Note, that the notation of the  "unitarity" triangle is introduced in the theory of the Weak Interactions in  accordance to the fact that the Cabibbo-Kobayashi-Maskawa~(CKM) matrix is an unitarity matrix (see e.g. \cite{Ceccucci:2008zz}). In our case the variables in Eq.~\ref{crossX} appear through the solution of the BFKL equation obtained with the use of the unitarity constraints~\cite{BLS2}.

\begin{figure}[htbp]
	\begin{center}
		\epsfig{figure=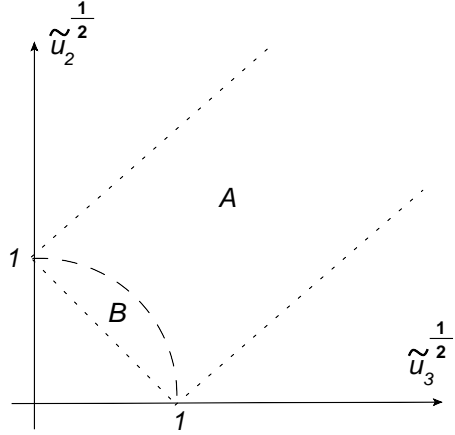,width=60mm}
	\end{center}
	\caption{The region of the reduced cross ratios where the analytic continuation is performed.   }
	\label{fig:region}
\end{figure}

We see from Fig.~\ref{fig:triangle} that the reduced cross ratios should obey the triangle inequalities
\beq\label{regionredu}
\sqrt{\tilde{u}_2}+\sqrt{\tilde{u}_3}\geq 1, \;\;\; 1+\sqrt{\tilde{u}_2} \geq \sqrt{\tilde{u}_3}, \;\;\;
1+\sqrt{\tilde{u}_3}\geq \sqrt{\tilde{u}_2}
\eeq
or in terms of $u_2$ and $u_3$
\beq \label{regionu}
\sqrt{u_2}+\sqrt{u_3}\geq \sqrt{1-u_1},\;\;\; \sqrt{u_3}-\sqrt{u_2}\leq \sqrt{1-u_1},\;\;\; \sqrt{u_2}-\sqrt{u_3}\leq
\sqrt{1-u_1}
\eeq
Note, that the parameter $\Delta$ appearing in Eq.~\ref{R6}  is proportional to the area of the "unitarity" triangle expressed by the Heron formula in terms of its sides.
As we have already mentioned we perform the analytic continuation in $u_1$ with an additional condition of the cross ratios $u_1+u_2+u_3<1$. This condition is needed solely for avoiding  possible difficulties in the continuation of the function $\chi$ in Eq.~\ref{chi}, and can be written as
\beq \label{extracond}
\tilde{u}_2+\tilde{u}_3<1
\eeq
The regions limited by the constraints Eq.~\ref{regionredu} and Eq.~\ref{extracond} are illustrated in Fig.~\ref{fig:region}. The region $\mathbf{A}$ is limited by Eq.~\ref{regionredu} and  corresponds to multi-Regge kinematics, while its subregion $\mathbf{B}$ is the region where the condition $u_1+u_2+u_2<1$ is valid. The remainder  function after the analytic continuation in the region $\mathbf{B}$ does not have any singularities on the boundary of Eq.~\ref{extracond} and thus it is valid in the whole region $\mathbf{A}$, including its boundaries.

Based on Eq.~\ref{Kvarcomplex} we introduce complex variables 
\begin{eqnarray}\label{paramComplex}
z=\sqrt{\tilde{u}_2}e^{i\phi_2},\;\;\;1-z=\sqrt{\tilde{u}_3}e^{-i\phi_3}.
\end{eqnarray}
as illustrated in Fig.~\ref{fig:triangle}.
The parametrization of Eq.~\ref{paramComplex} for an arbitrary complex $z$ is compatible with the constraints of the multi-Regge kinematics given by Eq.~\ref{regionredu} and allows to eliminate the square roots $\sqrt{\Delta}$ in the arguments.
In particular, using the complex variables of Eq.~\ref{paramComplex} we can show explicitly that a potentially dangerous line $\Delta=0$ ( see Eq.~\ref{Xpm} for the definition of $\Delta$) that could lead to singularities of the remainder function, causes no problems since all $x_{i}^{\pm}$ are replaced by  $(1-z)/z$ and its complex conjugate. 
It is worth emphasizing that we perform the analytic continuation of Eq.~\ref{R6} in variables $u_i$ and only then pass to the parametrization of Eq.~\ref{paramComplex}. This allows to avoid unnecessary difficulties related to the dependence of the variable $z$ on the cross ratio $u_1$.
The result of the analytic continuation is simplified leaving only leading  and the constant term in the logarithm of energy $\ln(1-u_1)\simeq -\ln s_2$. The terms that are suppressed by the power of $s_2$ are omitted in our calculations. At the intermediate steps of the analytic continuation there appeared contributions of the order of $\ln^2 (1-u_1)$ and $\ln^3(1-u_1)$, these terms are incompatible with the unitarity approach and fortunately all cancel in the final expression. Some terms also develop a real part proportional to $(-i2\pi)^2$ during the analytic continuation, but they all cancel out as well. The final result of the analytic continuation of Eq.~\ref{R6} in the multi-Regge kinematics is presented in  Eq.~\ref{KR62result}.

We have calculated the remainder  function $R(|u_1|e^{-i2\pi},|z|^2(1-u_1),|1-z|^2(1-u_1))$  in  Eq.~\ref{KR62result} under condition  $u_1+u_2+u_3<1$ (in the region $\mathbf{B}$ of Fig.~\ref{fig:region}), but the resulting expression does not have singularities on  the boundary, so it is valid also for $u_1+u_2+u_3\geq 1$ (in the whole  region $\mathbf{A}$ of Fig.~\ref{fig:region}).  Same is true also for condition $\Delta<0$ used for the analytic continuation; the resulting expression is valid for any value of $\Delta$. 

 The function  of Eq.~\ref{KR62result} has the  $\tilde{u}_2 \leftrightarrow \tilde{u}_3$~($z\leftrightarrow 1-z$) symmetry, which corresponds to the target-projectile symmetry (see Fig.~\ref{fig:6point}). Eq.~\ref{KR62result} vanishes if  either $z\to 1$ or $z\to 0$, which is the case when the momentum of one of the produced particles $k_i$ in Fig.~\ref{fig:6point} goes to zero.
The expression of Eq.~\ref{KR62result}  explicitly demonstrates that the remainder  function is pure imaginary in the multi-Regge kinematics, which means that non-analytic terms $\sqrt{\Delta}$ canceled out.
The first term on the RHS  of Eq.~\ref{KR62result} reproduces the LLA prediction of the BFKL approach found by one of the authors with collaborators~\cite{BLS2} and given by Eq.~\ref{KcorLLAreduced}, while the rest of the RHS in Eq.~\ref{KR62result} present the NLO contribution not yet calculated using BFKL formalism. This way we find an agreement between the BFKL approach to the Mandelstam cut contributions   and the calculations exploited the Wilson Loop/Scattering Amplitude duality, at least at the level of the leading logarithm of the energy.

\section{Conclusion}
In this paper we performed  an analytic continuation of the GSVV~\cite{Goncharov:2010jf} formula for the remainder function of two-loop six-point MHV amplitude in SYM $\mathcal{N}=4$ to the region, where all but two energy invariants are negative. The result is then simplified in the multi-Regge kinematics and is shown to agree with the calculations in the BFKL approach~\cite{BLS2}. In particular we reproduce the leading logarithmic  term and find the cancellation of the real part of the remainder function in this limit in an agreement with  predictions based on the analyticity and factorization~\cite{Lipatov:NEW}. We also extract subleading terms, which were not yet calculated in the BFKL formalism. These terms are pure imaginary and have correct analytic properties. This supports the validity of the relation between the light-like Wilson loops and the planar MHV scattering amplitudes in $\mathcal{N}=4$ SUSY for  a weak coupling constant.  The details of our calculations will be presented in the next paper~\cite{Prygarin:NEW}.

\section{Acknowledgments}
We deeply indebted to J.~Bartels for the  enlightening remarks and continued support.  
A.P. is grateful to G.~P.~Korchemsky for fruitful discussions and hospitality in Saclay, where part of this work has been done.

\end{document}